\documentclass[12pt]{article}
\usepackage{graphicx}
\newlength{\extraspace}
\newlength{\extraspaces}
\newcommand{\be}{\begin{equation}}
\newcommand{\ee}{\end{equation}}
\newcommand{\bea}{\begin{eqnarray}}
\newcommand{\nn}{\nonumber}
\newcommand{\eea}{\end{eqnarray}}

\def\lsim{\mathrel{\rlap {\raise.5ex\hbox{$ < $}}
{\lower.5ex\hbox{$\sim$}}}}

\newcommand{\pr}{\paragraph{}}
\def\gappeq{\mathrel{\rlap {\raise.5ex\hbox{$>$}}
{\lower.5ex\hbox{$\sim$}}}}
\def\lappeq{\mathrel{\rlap{\raise.5ex\hbox{$<$}}
{\lower.5ex\hbox{$\sim$}}}}

\begin{document}

\begin{titlepage}

\begin{flushright}
OUTP-99-41P \\
hep-th/9909018 
\end{flushright}

\begin{centering}
\vspace{.05in}
{\Large {\bf Infinitely Coloured Black Holes\\}}
 
\vspace{.4in}

{\bf {Nick E. Mavromatos$^{*}$ and Elizabeth Winstanley}} \\
\vspace{.05in}
University of Oxford, Department of Physics, Theoretical Physics,
1 Keble Road, Oxford, OX1 3NP, United Kingdom. \\
\vspace{.4in}
{\bf Abstract} \\
\vspace{.05in}
We formulate the field equations for $SU(\infty )$ 
Einstein-Yang-Mills theory, and find spherically symmetric
black-hole solutions.
This model may be motivated by string theory considerations,
given the enormous gauge symmetries which characterize
string theory.
The solutions simplify considerably in the presence of a negative
cosmological constant, particularly for the limiting
cases of a very large cosmological constant or very small gauge
field.
The situation of an arbitrarily small gauge field is 
relevant for holography and we comment on the AdS/CFT
conjecture in this light.
The black holes possess infinite amounts of gauge field hair,
and we speculate on possible consequences
of this for quantum decoherence, which,
however, we do not tackle here.
\end{centering}
\vspace{2.5in}
\vfill

\begin{flushleft} 
September 1999 \\ [.2in]
$^{*}$~Currently also at Theory Division, \
CERN, CH-1211 Geneva 23, Switzerland. \\ 
\end{flushleft} 

\end{titlepage}
\newpage
\section{Introduction}

\pr
The discovery of the Bartnik-McKinnon self-gravitating
particle-like structure~\cite{bartnik} has opened the 
way for the construction of several 
hairy black hole structures in 
Einstein theories coupled to non-Abelian 
gauge fields~\cite{colour}.
In this work we would like to address 
some issues concerning black hole solutions 
in $SU(\infty)$ gauge theories. 
Our interest in such structures 
will be motivated by string theory considerations. 
String theories are known to be characterized 
by infinite-dimensional gauge symmetries,
relating the various string levels. 
Such infinite symmetries have been argued~\cite{kalara,emn}
to be sufficient for the maintainance of quantum 
coherence during black hole (quantum) evaporation.
The basic idea behind such conjectures
is that infinite symmetries are sufficient 
to encode all the information inside a stringy  black hole,
since the latter can be in any one of an infinity 
of excited massive-string level  
states during the evaporation, thereby accounting for the 
enormous number of states that appears to characterize
the Bekenstein-Hawking entropy formula for macroscopic black holes. 
This obviously could not have been achieved in 
a local field theory with a finite number of conserved
charges. 
\pr
However, the point of view of ref. \cite{emn} 
was that simply infinite-dimensional symmetries
were not by themselves sufficient to maintain coherence,
during the complex process of evaporation/decay in string theory.
Studies in prototypes of two-target-space-dimensional 
string theories
have revealed that the latter possess black hole structures
which are characterized~\cite{emn,bakas} 
by specific infinite-dimensional 
symmetries, namely $W_\infty$ symmetries~\cite{bakas}, in their target 
space. 
Such symmetries have the {\it peculiar} property of preserving 
a two-dimensional phase-space manifold under target-time translations,
which in the particular case of two-dimensional 
strings may be identified with the 
phase space of scalar massless matter in the two-dimensional 
black hole space time. Hence, during black-hole quantum evaporation,
which according to ref. \cite{emn}, is a non-thermal phenomenon
associated with the decay 
of excited black hole (internal) string states, 
quantum coherence of the string matter is maintained as a result 
of {\it geometrical properties of the black hole 
hair itself ($W$-hair)}. 
\pr
Although this phenomenon seems to be a novel
way of tackling the coherence issue, however, 
the fact that in ref. \cite{emn} has been explicitly demonstrated 
in two-dimensional black holes has the drawback that 
it might be connected strictly to the two-dimensional 
space time. Such objections can be answered, of course, 
by making an analogy of the two-dimensional 
stringy black hole with an effective theory 
arising in realistic four-dimensional theories 
of quantum gravity upon restricting oneself to 
spherically-symmetric space-time configurations;
but then the challenge remains to construct explicitly 
string theories in four dimensions with this property. 
\pr
Motivated by these considerations, we would like in this 
note to take a different viewpoint, and 
consider black holes 
in standard Einstein four-dimensional 
space-times, which however
capture {\it some} of the essential 
features of the above-mentioned
two-dimensional stringy black holes, namely 
the {\it infinite-dimensional} character of 
the gauge hair, as well as its area-preserving 
nature, which was argued to be responsible for 
two-dimensional quantum coherence~\cite{emn}.
At this point we should stress  that 
the $W$-hair of ref. \cite{emn} was {\it quantum} hair~\cite{qhair},
arguably measured by Aharonov-Bohm type experiments,
and therefore pertaining to phases in the respective 
matter wave functions, 
whereas in this paper we are concerned with purely
{\it classical} hair.
Nevertheless, some of the features we find, specifically
the holographic properties, may be related to the issue
of quantum coherence in the sense of ref.~\cite{emn}.
\pr
Gauge theories  with gauge group $SU(N)$, 
$N \rightarrow \infty$, remarkably share
these features, since gauge transformations in the large-$N$ limit
yield an ancestor of the $W_\infty$ algebra, according to the 
analysis of ref. \cite{floratos}. Indeed, in this limit 
the non-Abelian commutators of 
such gauge transformations become {\it classical-like} Poisson 
brackets, which have the property of {\it preserving the area }
of an ``internal'' sphere, $S^2$, pertaining to the 
gauge group variables. For details on this construction we refer the
reader
to the literature~\cite{floratos}; here we shall make use 
only of the results relevant for our purposes, namely 
the technicalities involved in taking the limit 
$N \rightarrow \infty$. 
Black hole solutions of the Einstein-Yang-Mills (EYM) 
equations for finite $N$ have been discussed in 
\cite{jmp}. However, as was pointed out there,
the finite-$N$ analysis could not be extended smoothly 
to the case $N \rightarrow \infty$.
It is the purpose of this article to establish the 
existence of 
$SU(\infty)$ {\it classical} gauge hair in the EYM
black-hole system.  
\pr
The area-preserving nature of the infinite-$N$ gauge hair
is an encouraging sign for the maintainance of quantum 
coherence during the evaporation process; however, to prove 
this conjecture one should study the precise mechanism
behind the evaporation, which in view of the absence
of a satisfactory four-dimensional quantum theory of gravity, 
is not feasible at present.
Nevertheless, as we shall show in this article, 
the analytic 
formulation of black hole solutions with infinite
gauge hair simplifies greatly in the presence of a
negative cosmological constant (no matter how small).
We have arbitrarily small gauge field hair (which is not
possible in the absence of a cosmological constant).
This implies that anti-de Sitter geometries may be used
as ``regulators'' for the $SU(\infty )$ EYM black holes
with zero cosmological constant,
and therefore one can appeal to the concept of holography
in the sense of \cite{thooftsussk}.
This means that the behaviour of the gauge field in the
bulk (i.e. outside the event horizon) is determined
by the structure at infinity (boundary). 
Anti-de Sitter space-times have recently attracted great attention
in view of their relevance to the conjectured equivalence
between some conformal field theories on their boundary
with the classical bulk structure 
(AdS/CFT correspondence)~\cite{maldacena,witten}.
The above-mentioned holographic point of view, together
with the specific area-preserving properties of the 
$SU(\infty )$ gauge group, makes this model an
interesting classical hair analogue of the $W$-hair picture
of \cite{emn}.
Whether it survives at the full quantum level remains an
open question.
\pr
The structure of the present paper is as follows. 
In section 2, we review briefly the mathematical 
formalism behind Einstein-Yang-Mills systems, 
by concentrating on the construction of the field equations 
for spherically symmetric black hole solutions for $SU(\infty)$
gauge group. 
We take carefully the limit $N \rightarrow \infty$,
by following the same procedure as ref. \cite{floratos},
which led to the area-preserving property of the $SU(\infty)$ 
gauge transformations. 
We include a cosmological constant $\Lambda$ in our model.
In section 3 we consider the limit in which the cosmological
constant is large and negative, since in this limit it is 
known that the $SU(2)$ system simplifies significantly~\cite{eliz}.
We perform an asymptotic expansion in $\epsilon =1/\Lambda $
and obtain analytic solutions to first order in $\epsilon $,
which reveal many of the properties we expect the black holes
to possess in general.
We also consider the solutions for a very small gauge field, 
when the field equations again simplify considerably.
As already mentioned,
this situation is of particular interest for our 
ideas concerning holography~\cite{thooftsussk} and
the AdS/CFT correspondence~\cite{maldacena,witten}.

\section{Ansatz and field equations for $SU(\infty)$ Gauge Field}
\subsection{Mathematical Construction of the Gauge Field}
\pr
It is the purpose of this section to study in detail 
the mathematical construction of an $SU(N)$ gauge field,
extending in a non-trivial way the flat-space-time approach
of ref.~\cite{floratos}.
First we shall keep $N$ finite, and at the end we shall take 
the limit $N \rightarrow \infty$ in a similar way as the one 
considered in \cite{floratos}. 
The mathematical issue of how $SU(\infty )$ is
constructed is rather subtle, but we shall
not focus on these difficulties.
Instead, we shall assume that the construction
in \cite{floratos} is well-defined and
take that as our definition of $SU(\infty )$.
We restrict our attention to space-times which are 
static  and spherically 
symmetric, and shall therefore use the analysis
of the spherically symmetric $SU(N)$ gauge field \cite{kunzle}.
Throughout this paper, the metric has signature $(-,+,+,+)$.
We shall retain the gravitational coupling constant $\kappa $
but set $c=1$.

\pr
Following \cite{floratos,dewit} one starts from 
an irreducible representation of the 
$su(N)$ Lie algebra, spanned by $N \times N$ Hermitian 
matrices $S_i$:
\bea
S_1 & = & 
\frac{1}{2}
\left(
\begin{array}{cccccc}
0 & \sqrt{N-1} & 0 &  \dots & 0  & 0 \\
\sqrt{N-1} & 0 & \sqrt{2(N-2)} &  \dots  & 0 & 0 \\
0 & \sqrt{2(N-2)} & 0 & \dots  & 0 & 0\\
\vdots & \vdots & \vdots & \vdots & \vdots & \vdots \\
0 & 0 & 0 & \dots & 0 & \sqrt{N-1} \\
0 & 0 &  0 & \dots & \sqrt{N-1} & 0 
\end{array} \right)
\nn \\
S_2 &  =  & 
\frac{1}{2}
\left(
\begin{array}{cccccc}
0 & -i\sqrt{N-1} & 0 & \dots & 0 & 0 \\
i\sqrt{N-1} & 0 & -i\sqrt{2(N-2)} & \dots & 0 & 0\\
0 & i\sqrt{2(N-2)} & 0 & \dots & 0 & 0  \\
\vdots & \vdots & \vdots & \vdots & \vdots & \vdots  \\
0 & 0 & 0 & \dots & 0 &  -i\sqrt{N-1} \\
0 & 0 & 0 & \dots &  i\sqrt{N-1} &  0 
\end{array}\right) 
\nn \\
S_3 & =  &
\frac{1}{2}
\left(
\begin{array}{ccccc} 
N-1 & 0 & 0 & \dots & 0 \\
0 & N-3 & 0 & \dots & 0 \\
0 & 0 & N-5 & \dots & 0 \\
\vdots & \vdots & \vdots & \vdots & \vdots   \\
0 & 0 & 0 & \dots &  -N+1 
\end{array}\right) 
\label{s123}
\eea
which satisfy the commutation relations
\be
[S_i, S_j ]=i\epsilon_{ijk}S_k.
\label{commrel}
\ee
From (\ref{s123}) one can construct the `raising' and `lowering' 
matrices:
\bea
S_+  \equiv S_1 + i S_2 
& = & 
\left(
\begin{array}{cccccc}
0 & \sqrt{N-1} & 0 & 0 & \dots & 0  \\
0 & 0 & \sqrt{2(N-2)} &  0 & \dots & 0 \\
0 & 0 & 0 & \sqrt{3(N-3)} & \dots & 0 \\
\vdots & \vdots & \vdots & \vdots & \vdots & \vdots  \\
0 & 0 & 0 & 0 &  \dots &  \sqrt{N-1} \\
0 & 0 & 0 & 0 & \dots & 0 
\end{array}\right) 
\nn \\
S_-  \equiv S_1 - i S_2 
& = & 
\left(
\begin{array}{cccccc}
0 & 0 & 0 & \dots & 0 & 0\\ 
\sqrt{N-1} & 0 & 0 & \dots & 0 & 0  \\
0 & \sqrt{2(N-2)} &  0 & \dots &  0 & 0\\
\vdots & \vdots & \vdots & \vdots & \vdots & \vdots   \\
0 & 0 &  0 &\dots & \sqrt{N-1} & 0
\end{array}\right) . 
\label{spm}
\eea
The most general spherically symmetric 
$SU(N)$ gauge potential may be written in the differential 
form (where all matrices are functions of the radial
Schwarzschild co-ordinate $r$)~\cite{kunzle}
\be
{\hat {\cal A}}= A^{(N)}dt + B^{(N)} dr + \frac{1}{2}(C^{(N)}-C^
{(N)\dagger})d\theta 
-\frac{i}{2}[(C^{(N)}+C^{(N)\dagger})
 \sin \theta + D^{(N)}\cos \theta ]d\phi 
\label{potential}
\ee
where $D^{(N)}={\rm {Diag}}\{k_1, \dots ,k_N\}$, with 
$k_1 \ge k_2 \ge \dots \ge k_N$
integers whose sum is {\it zero}. 
The $N\times N$ matrix $C$ is  strictly upper 
triangular, with  complex entries satisfying 
$C_{ij}^{(N)} \ne 0$ only if $k_i=k_j + 2$; and $C^\dagger$ 
is its Hermitian conjugate.  
The anti-Hermitian matrices $A$, $B$  commute with $D$.
In addition, $D$ must satisfy the commutation relations:
\be
[D,C]=2C, \qquad [D,C^{\dagger }]=-2C^{\dagger }.
\label{dconds}
\ee  
A suitable irreducible ansatz is to take~\cite{kunzle}:
\be
D^{(N)}={\rm {Diag}}\{N-1, N-3, \dots -N+3, -N+1 \} .
\label{ansatzD}
\ee
The matrices $A,B$ are now diagonal with trace zero, and
their entries can be written as:
\be
A_{jj}^{(N)} =i
\left[ -\frac{1}{N} \sum_{k=1}^{j-1}k{\cal A}_k^{(N)} 
+ \sum_{k=j}^{N-1}
\left( 1-\frac{k}{N} \right) {\cal A}_k^{(N)} 
\right] ;
\label{AB}
\ee
and similarly for $B^{(N)}$. 
Here ${\cal A}_k^{(N)}$ are real functions. The only non-vanishing 
entries for $C^{(N)}$ are:
\be
C_{j,j+1}^{(N)}=\omega_je^{i\gamma_j}, \qquad j=1, \dots N-1 
\label{C}
\ee
for real functions $\omega_j, \gamma _j$ depending on $r$. 
In this case, $D=2S_3$, and 
\be
 A^{(N)}=a_0I + a_1S_3 + \dots + a_{N-1}S_3^{N-1} 
\label{Aexp}
\ee
with a similar expression for $B^{(N)}$. The coefficients $a_i(r)$ 
are purely imaginary. In addition,
\be
C^{(N)}=c_0S_+ + c_1[S_+S_3 + S_3S_+] + \dots + c_{N-2} [S_+S_3^{N-2} 
+ S_3S_+S_3^{N-3} + \dots S_3^{N-2}S_+]
\label{Cexp}
\ee
where the $c_i(r)$ are complex numbers, so that 
\be
C^{(N)\dagger}=c_0^*S_- + c_1^*[S_-S_3 + S_3S_-] + \dots +
c_{N-2}^*[S_-S_3^{N-2} 
+ \dots S_3^{N-2}S_-] .
\label{Cstarexp}
\ee

\pr
The next stage is to rewrite this ansatz in a form suitable for
taking the limit $N\rightarrow \infty $.
Following \cite{floratos} we now write the matrices $A,B,C$ and $D$
in terms of the matrix polynomials
\be
 {\tilde T}_{l,m}^{(N)} = \sum_{i_k=1,2,3} a_{i_1 \dots
i_l}^{(m)}S_{i_1}\dots
S_{i_l} 
\label{tmatrix}
\ee
for $l=1, \dots N-1$, $m=-l, \dots l$, where $a^{(m)}_{i_1 \dots i_l}$
is a symmetric and traceless tensor given by the spherical 
harmonics:
\be
  Y_{l,m}(\theta, \phi) =
\sum_{
\begin{array}{c}
i_k=1,2,3; \\ 
k=1, \dots l 
\end{array}}
a_{i_1\dots i_l}^{(m)}\eta_{i_1} \dots \eta_{i_l}
\label{harmonics}
\ee
where
$\eta _1= \cos \phi  \sin \theta $, 
$\eta_2 = \sin \phi \sin \theta$, 
$\eta_3=\cos \theta$.
Using the variables $\eta_+ = e^{i\phi}\sin \theta 
= \eta_1 + i\eta_2 $, $\eta_-=e^{-i\phi}\sin \theta = \eta_1
-i\eta_2$, the relevant spherical harmonics are:
\bea
Y_{l,0}(\theta, \phi) & = & 
\left[ \frac{2l +1}{4\pi} \right] ^{\frac{1}{2}}
P_l(\cos \theta)
\nn \\
Y_{l,1}(\theta, \phi) & = & 
-\left[ \frac{2l+1}{4\pi}\frac{(l-1)!}{(l+1)!} \right] ^{\frac{1}{2}}
P_l^{1}(\cos \theta)e^{i\phi} \nn \\
Y_{l,-1}(\theta, \phi) & = & 
-\left[ \frac{2l+1}{4\pi}\frac{(l-1)!}{(l+1)!} \right]^{\frac{1}{2}}
P_l^{1}(\cos \theta)e^{-i\phi} ;
\label{harmonics2}
\eea
where
\be
P_l(\cos \theta)=\frac{1}{2^l l!}
\left(\frac{d}{d(\cos \theta)}\right)^l(\cos ^2\theta -1)^l
\ee 
is a Legendre polynomial, and 
\be
P_l^{1}(\cos \theta) = 
(1-\cos ^2\theta)^{\frac{1}{2}}\left(\frac{d}{d(\cos \theta )}
\right)P_l(\cos \theta)
\ee  
is an associated Legendre function. 
Since $P_l(\cos \theta)$ is a polynomial of order $l$, 
then $P_l^{1}=\sin \theta \times {\rm a~polynomial~of~order~}l-1$, 
so that 
\bea
Y_{l,0}(\theta, \phi)& \propto & P_l(\eta_3) \nn \\
Y_{l,1}(\theta, \phi)& \propto & \eta_+P_l'(\eta_3) \nn \\
Y_{l,-1}(\theta, \phi)& \propto  &\eta_-P_l'(\eta_3)  .
\label{moreharm}
\eea
The spherical harmonics in this form are not homogeneous
(as required by (\ref{harmonics})), 
however they can be made homogeneous by including 
suitable powers of $\eta_1^2 + \eta_2^2 + \eta_3^2 =1$.

\pr
The ansatz is then
\bea
A^{(N)} &  = & 
{\tilde a}_0^N{\tilde T}_{0,0}^{(N)} + {\tilde a}_1^N 
{\tilde T}_{1,0}^{(N)} + \dots + {\tilde a}_{N-1}^N {\tilde
T}_{N-1,0}^{(N)} 
\nn \\
B^{(N)} &  = &  {\tilde b}_0^N{\tilde T}_{0,0}^{(N)} + {\tilde b}_1^N 
{\tilde T}_{1,0}^{(N)} + \dots + {\tilde b}_{N-1}^N {\tilde
T}_{N-1,0}^{(N)} 
\nn \\
C^{(N)} & = & {\tilde c}_1^N{\tilde T}_{1,1}^{(N)} + {\tilde c}_2^N 
{\tilde T}_{2,1}^{(N)} + \dots + {\tilde c}_{N-1}^N {\tilde
T}_{N-1,1}^{(N)} 
\nn \\
D^{(N)} & = & {\tilde d}_{1}{\tilde T}_{1,0}^{(N)} .
\label{abcd}
\eea
As before, ${\tilde a}_i, {\tilde b}_i$ are purely imaginary, 
${\tilde c}_i$ are complex, and ${\tilde d}_{1}$ is real. 
Due to the self-adjointness of $S_i$, $i=1,2,3$. 
the ${\tilde T}$ satisfy $[{\tilde
T}_{l,m}^{(N)}]^\dagger=(-1)^m{\tilde {T}}_{l,-m}^{(N)}$. 
Hence, 
\be
   C^{(N)\dagger} =-({\tilde c}_1^{(N)})^*{\tilde T}_{1,-1}^{(N)} - \dots
-({\tilde c}_{N-1}^{(N)})^*{\tilde T}_{N-1,-1}^{(N)} .
\label{ctildahermitean}
\ee

\pr
In order for the theory to have a well-defined limit as $N \rightarrow
\infty$, one should form a new basis~\cite{floratos,dewit}
for the $SU(N)$ group, $\{ T_{l,m}^{(N)} \}$, by replacing 
$S_i$ in ${\tilde T}_{l,m}^{(N)}$ by $\tau_i$ defined by:
\be
\tau _i \equiv \frac{1}{N}S_i ,
\label{taui}
\ee
so that the ansatz (\ref{abcd}) now has the form 
\be
A^{(N)} = a_0^{(N)} T_{0,0}^{(N)} + a_1 T_{1,0}^{(N)} 
+ \dots a_{N-1}^{(N)}T_{N-1,0}^{(N)}
\label{newa}
\ee
and similarly for the other matrices. 
Note that $A^{(N)}$ as defined here corresponds to a rescaled
gauge potential, as compared with that 
in ref. \cite{kunzle}, by a factor  
$\frac{1}{N}$; this is due to the fact that $A^{(N)}$ has a finite
limit 
as $N \rightarrow \infty$. 
We are now in a position simply to write down the ansatz 
for the $SU(\infty )$ gauge field.
The result is:
\be
A  = 
\sum _{l=1}^{\infty } a_{l} Y_{l,0}
\qquad
B  =  
\sum _{l=1}^{\infty } b_{l} Y_{l,0}
\qquad
C  =  
\sum _{l=1}^{\infty } c_{l} Y_{l,1}
\qquad
D  =  
d_{1} Y_{1,0} ,
\label{infans}
\ee 
where $a_{l},b_{l}$ and $c_{l}$ depend on $r$, 
and $d_{1}$ is a real constant.

\subsection{Field Equations for Einstein-SU($\infty$) 
Yang-Mills Theory}

\pr
The Einstein and Yang-Mills equations for an $SU(N)$ 
gauge theory, in a spherically-symmetric space-time,  
have been derived in ref. \cite{kunzle}. 
It is the point of this subsection to 
re-derive these equations in the limit $N \rightarrow \infty$. 
To this end we first start by briefly 
reviewing the results of ref. \cite{kunzle}, that are relevant 
for our purposes here. For more details on the finite-$N$ $SU(N)$ gauge
theory
we refer the interested reader
to ref. \cite{kunzle}. 

\pr
Firstly, we consider the Yang-Mills equations.
One component of these leads to the following equation, for the
ansatz (\ref{C}), where none of the $\omega _{j}^{(N)}$ vanish,
\be
{\cal B}_j^{(N)} + \gamma_j^{'(N)} =0 .
\label{eqs2}
\ee 
So far we have not fixed our choice of gauge, so we use the remaining
gauge freedom we have to set ${\cal {B}}_{j}^{(N)}=0$ for static solutions
\cite{brodbeck}, and then on account of (\ref{eqs2}) one has
$\gamma _{j}^{'(N)}=0$.
This implies that each $\gamma _{j}^{(N)}$ is a constant, which we
set equal to $0$ for simplicity.
In addition, we shall concentrate on purely magnetic field 
configurations by setting ${\cal {A}}_{j}^{(N)}=0$ for all $j$.

\pr
In order to obtain a finite limit of the field equations as
$N\rightarrow \infty $, care is needed with factors of $N$.
The ansatz used in the equations below is (\ref{newa}), with the
basis of $SU(N)$ matrices constructed from the rescaled matrices 
$\tau _{i}$ (\ref{taui}).
The field equations can then be taken
from \cite{kunzle}, making the following
substitution  \cite{floratos}
\be
{\hat {\cal {A}}} \rightarrow N {\hat {\cal {A}}} .
\label{ntimes}
\ee
We consider a spherically symmetric geometry described by 
the following metric, in Schwarzschild co-ordinates:
\be
ds^{2}=-\mu e^{2\delta (r)} \, dt^{2} +\mu ^{-1} dr^{2} + r^{2}
\left( d\vartheta ^{2} 
+\sin ^{2} \vartheta \, d\varphi ^{2} \right) ,
\ee
where the metric function $\mu $ has the form
\be
\mu = 1  -\frac {2m(r)}{r} -\frac {\Lambda r^{2}}{3} ,
\ee
and we have introduced a cosmological constant $\Lambda $.
It should be emphasised at this point that the co-ordinates
$\vartheta $, $\varphi $ are different from the 
``internal'' variables $\theta $, $\phi $, which appeared in the
ansatz for the $SU(\infty )$ gauge fields in the previous section.
For finite $N$, the remaining Yang-Mills equation is,
with $'$ denoting derivative with respect to $r$,
\be
r^{2} \mu C''+
\left( 2m -\frac {2\Lambda r^{3}}{3} -\kappa r^{3} p_{\vartheta }
\right) C' +C +\frac {1}{2} N^{2} [C,[C,C^{\dagger }]] =0
\label{YM}
\ee
where $\kappa $ is the gravitational coupling constant
(we have set the gauge field coupling constant to unity), and 
\be
p_{\vartheta }= \frac {1}{r^{4}} {\rm {Tr}} 
\, (D-N[C,C^{\dagger }] )^{2} .
\ee
The Einstein equations read
\be
m' =  \frac {\kappa }{2} \left(
 \mu G +r^{2} p_{\vartheta }  \right) ,
\qquad
\delta ' = \frac {\kappa }{r} G
\label{einstein}
\ee
where
\be
G={\rm {Tr}} \, (C'C^{'\dagger }) .
\ee
The presence of the explicit factors of $N$ is due to the 
substitution (\ref{ntimes}) and will be essential for
obtaining the limit $N\rightarrow \infty $.

\pr
In taking the limit $N\rightarrow \infty $, the matrix $C$ becomes
a function of the internal co-ordinates
$\theta, \phi $ (i.e. those variables which are the arguments
of the spherical harmonics, rather than the space-time 
co-ordinates $\vartheta $ and $\varphi $). 
The following substitutions are made \cite{floratos}:
\bea
{\rm {Tr}} & \rightarrow & \frac {1}{4\pi }  \int _{\theta =0}^{\pi }
\int _{\phi =0}^{2\pi } \sin \theta \, d\theta \, d\phi 
\nn 
\\
N[P , Q ] & \rightarrow & i\left\{ P(r,\theta ,\phi ), Q(r,\theta ,
\phi ) \right\} ,
\label{nlimits}
\eea
where the Poisson bracket is defined as:
\be
\left\{ P, Q \right\}
= \frac {\partial P}{\partial (\cos \theta )}
\frac {\partial Q}{\partial \phi }
-\frac {\partial P}{\partial \phi }
\frac {\partial Q}{\partial (\cos \theta )}.
\ee
The Yang-Mills equation (\ref{YM}) therefore takes the form
\be
r^{2} \mu 
\frac {\partial ^{2}C}{\partial r^{2}}+
\left( 2m -\frac {2\Lambda r^{3}}{3} -\kappa r^{3} p_{\vartheta }
\right) \frac {\partial C}{\partial r} +C 
-\frac {1}{2}  \{ C, \{C,C^{\dagger }\} \} =0
\label{infYM}
\ee
while the Einstein equations retain the form (\ref{einstein}), with
\bea
G & = & 
\frac {1}{4\pi } \int _{\theta =0}^{\pi }
\int _{\phi =0}^{2\pi } 
\frac {\partial C}{\partial r} 
\frac {\partial C^{\dagger }}{\partial r} 
\sin \theta \, d\theta \, d\phi 
\nn
\\
p_{\vartheta } & = & 
\frac {1}{16\pi r^{4}} \int _{\theta =0}^{\pi }
\int _{\phi =0}^{2\pi } 
\left[ D -i \{ C, C^{\dagger } \} \right] ^{2}
\sin \theta \, d\theta \, d\phi .
\label{gravstuff}
\eea
The ansatz of the previous section (\ref{infans}) can be
written as:
\be
C(r,\theta, \phi )=f(r,\theta ) e^{i\phi }, 
\qquad
D= d_{1}\cos \theta ,
\ee
where $f$ is a real function and $d_{1}$ a real constant.
The generalization of the K\"unzle \cite{kunzle} 
ansatz for $SU(N)$
(\ref{ansatzD}) would imply that $d_{1}=2$.
However, in order for $D$ to satisfy the $SU(\infty )$ equivalent
of the conditions (\ref{dconds}), namely,
\be
i\{ D, C \} =2C, \qquad
i\{ D, C^{\dagger } \} = -2C^{\dagger }
\ee
it must be the case that $d_{1}=-2$.
We shall take this value of $d_{1}$ when we consider solutions of 
the field equations in the next section.

\pr
In terms of the co-ordinate $\xi =\cos \theta $, 
the Yang-Mills equation can be written as:
\be
r^{2}\mu \frac {\partial ^{2}f}{\partial r^{2}} 
+\left( 2m -\frac {2\Lambda r^{3}}{3} -\kappa r^{3}p_{\vartheta }
\right) \frac {\partial f}{\partial r} +f
+\frac {1}{2} f \frac {\partial ^{2}}{\partial \xi ^{2}} (f^{2}) =0;
\label{YMfinal}
\ee
and the quantities (\ref{gravstuff}) simplify to:
\bea
G & = & \frac {1}{2} \int _{\xi =-1}^{1} 
\left( \frac {\partial f}{\partial r} \right) ^{2} d\xi 
\nn
\\
p_{\vartheta } & = & \frac {1}{8r^{4}} 
\int _{\xi =-1}^{1} \left(
d_{1}\xi -\frac {\partial }{\partial \xi } 
(f^{2}) \right) ^{2} d\xi .
\label{gravfinal}
\eea
This is the simplest form of the equations for our purposes.
The function $f$ is further restricted by the ansatz (\ref{infans})
to be of the form
\be
f(r,\xi )= {\sqrt {1-\xi ^{2}}}
\sum _{i=1}^{\infty }f_{i}(r) \xi ^{i}.
\label{fform}
\ee
In the next section we shall consider black hole
solutions of the equations
(\ref{einstein},\ref{YMfinal},\ref{gravfinal}).

\section{Black hole solutions of the $SU(\infty )$ field equations}

\pr
We now turn to the solutions of the field equations 
(\ref{einstein},\ref{YMfinal},\ref{gravfinal}), 
focusing our attention on black hole geometries.
These equations are rather complicated, since they involve a
non-linear partial 
differential equation and integro-differential equations.
Given the fact that the $SU(N)$ equations have only been solved numerically,
tackling the field equations we have here would require a considerable
amount of complex numerical calculations.
The approach we shall take here is firstly to consider the special
cases, namely the embedded Schwarzschild, Reissner-Nordstr\"om and
$SU(2)$ black holes, in order to check our formalism and
fix the remaining constants.
Then we shall employ an asymptotic expansion in $\epsilon =1/\Lambda $
for the case $\Lambda <0$, $|\Lambda |\rightarrow \infty $.
This technique will enable us to find analytic, perturbative solutions
which, it is hoped, will have many of the physical properties
of the more general exact solutions but will be more amenable 
to analysis.
We are looking for black hole solutions with a regular event
horizon at $r=r_{h}$, and we want the geometry to 
approach (the covering space of)
anti-de Sitter space (AdS) as $r\rightarrow \infty $.
In the final subsection we shall make the alternative
approximation of a very small gauge field ($|f|\ll 1$),
in which case the field equations again simplify 
considerably, for all negative values of $\Lambda $.
This situation is particularly relevant in view of holography and
the AdS/CFT correspondence~\cite{maldacena,witten}.

\subsection{Embedded $SU(2)$ solutions}

\pr
Before discussing the genuine $SU(\infty )$ solutions of the field
equations (\ref{einstein},\ref{YMfinal},\ref{gravfinal}), 
we first consider embedded $SU(2)$ solutions.
For all $N$, it is known that $SU(N)$ EYM theory has neutral embedded
$SU(2)$ solutions  as well as charged embedded $SU(k)$ solutions
for $k=1,\dots , N-1$ \cite{jmp}.
The neutral $SU(2)$ solutions can be readily embedded into $SU(\infty )$
EYM.
However, this is not possible for the charged solutions, whatever
the value of $k$, including $k=2$.

\pr
The Yang-Mills function $f(r,\xi )$ is taken to be
\be
f(r,\xi )=w(r) {\sqrt {1-\xi ^{2}}} .
\label{fsu2}
\ee
Then the Yang-Mills equation (\ref{YMfinal}) reduces to:
\be
r^{2} \mu w'' +
\left( 2m -\frac {2\Lambda r^{3}}{3} -\kappa r^{3} p_{\vartheta }
\right) w' + w (1-w^{2}) =0,
\ee
and the quantities in the Einstein equations are:
\be
G=\frac {2}{3} w^{'2} , \qquad
p_{\vartheta }= \frac {1}{3r^{4}} (1-w^{2})^{2} ,
\ee
where we have fixed $d=-2$ as discussed in the previous subsection.
These equations are exactly the same as those for $SU(2)$ EYM
if we set $\kappa =3$.
This value of $\kappa $ is surprising, since it is conventional
in studies of $SU(N)$ EYM (see, for example \cite{jmp}) to 
take $\kappa =2$. 
The factor of $3$ arises from integrating the ${\sqrt {1-\xi ^{2}}}$
term, which does not arise in $SU(2)$ EYM.
In the following we shall take $\kappa =3$ in order that we can
check our results by comparison with the $SU(2)$ case.

\pr
There are two further special cases of the $SU(2)$ embedded solutions
which should be mentioned.
Firstly, setting $\omega \equiv \pm 1$ gives the 
Schwarzschild-AdS geometry, whilst setting $\omega \equiv 0$
gives a non-extremal
Reissner-Nordstr\"om-AdS black hole with charge
$Q=1$.

\subsection{Solutions for $|\Lambda |\gg 1$}

\pr
We now consider genuine $SU(\infty )$ solutions of the field equations
(\ref{einstein},\ref{YMfinal},\ref{gravfinal}).
In order to generate analytic solutions, we shall consider the
case where $\Lambda <0$, $|\Lambda |\gg 1$.
For $SU(2)$ EYM theory, the solutions are particularly simple in this
limit, and have attractive properties \cite{eliz}.
In particular, there are solutions which are linearly stable.
This limit corresponds to large black holes, which possess a stable
Hartle-Hawking state \cite{hp}. 
Therefore, these are the objects we are particularly interested in
if one wishes to consider issues of information loss during the evaporation
process. Also, in this limit stringy corrections in the bulk
geometry of AdS are negligible~\cite{maldacena,witten}, 
and the field theory limit is sufficient. 
\pr
Firstly, define a new parameter $\epsilon $ and a new variable $q$ by
\be
\epsilon = \frac {1}{\Lambda }, \qquad
q(r)= \frac {m(r)}{\Lambda },
\ee
where the variable $q$, although it is a metric function, will
be negative because $\Lambda <0$.
Then the field equations (\ref{einstein},\ref{YMfinal},\ref{gravfinal}) 
take the form
\bea
q'(r) & = &
\frac {3}{2} \left( \epsilon -\frac {2q}{r} -\frac {r^{2}}{3}
\right) G +\frac {3}{2} r^{2} \epsilon p_{\vartheta }
\nn
\\
\delta ' (r) & =  & \frac {3G}{r}
\nn
\\
0 & = & 
r^{2} \left( \epsilon -\frac {2q}{r} -\frac {r^{2}}{3} \right)
\frac {\partial ^{2}f}{\partial r^{2}} +
\left( 2q -\frac {2r^{3}}{3}- 3\epsilon r^{3} p_{\vartheta }
\right) \frac {\partial f}{\partial r}
+\epsilon f +\frac {1}{2} \epsilon f 
\frac {\partial ^{2}}{\partial \xi ^{2}} (f^{2}) ,
\eea
where we have set the gravitational coupling constant $\kappa =3$
in order to obtain the correct equations for the $SU(2)$ embedding,
as discussed in the previous subsection.
We shall consider solutions for $|\epsilon |\ll 1$, and 
perform an asymptotic expansion for small $\epsilon $ as follows:
\bea
q(r) & = & 
-\frac {r_{h}^{3}}{6} +\epsilon q_{1}(r)+\epsilon ^{2}q_{2}(r)
+\dots 
\nn
\\
\delta (r) & = & 1+ 
\epsilon \delta _{1}(r) +\epsilon ^{2} \delta _{2}(r) +\dots
\nn
\\
f(r,\xi ) & = & 
f_{0}(\xi ) +\epsilon f_{1}(r,\xi )+ \epsilon ^{2} f_{2} (r, \xi ).
+\dots
\label{asympt}
\eea
When $\epsilon =0$, the solution is a Schwarzschild-AdS black hole
with horizon radius $r_{h}$, and the gauge field $f$ is an arbitrary
function of $\xi $ of the form (\ref{fform}).
Here we are using an {\em {asymptotic expansion}}, in other words
the first few terms of the expansion above should be an increasingly
good approximation to the exact solutions as 
$\epsilon \rightarrow 0$, and the number of terms
of the expansion which are useful will also increase as
$\epsilon $ decreases.
This approximation will be reliable as long as the terms
in the expansions remain uniformly bounded in $r$.
This is a strong assumption, but one which is valid for the
terms in the expansion we obtain here.
The advantage of the expansion is that each term 
in the metric functions
$q$ and $\delta $ depends only on the {\em {previous}} 
terms in the expansion 
of the gauge function $f$, and for each term in $f$, we have a 
differential equation involving only derivatives with respect to $r$
for that term.
Therefore it is comparatively straightforward to find solutions in the
form (\ref{asympt}) to whatever order in $\epsilon $ we like. 
We shall find that interesting effects occur if we work to second order
in the metric functions and first order in the gauge field function.

\pr
To first order in $\epsilon $, we have:
\bea
\delta _{1}'(r) & = & 0 
\nn
\\
q_{1}'(r) & = & 
\frac {3C_{1}}{4r^{2}}
\nn
\\
{\cal {F}}_{1}(\xi ) & = &
r^{2} \left( r^{2}-\frac {r_{h}^{3}}{r} \right)
\frac {\partial ^{2}f_{1}}{\partial r^{2}} +
\left( r_{h}^{3}+2r^{3} \right) 
\frac {\partial f_{1}}{\partial r} 
\label{pert1}
\eea
where $C_{1}$ is a constant, and ${\cal {F}}_{1}$ a function of $\xi $,
given by
\be 
C_{1}= \int _{\xi =-1}^{1} \left[
\xi +\frac {1}{2} \frac {\partial }{\partial \xi } (f_{0}^{2})
\right] ^{2} \, d\xi ,
\qquad
{\cal {F}}_{1} =
3 \left[ f_{0} +\frac {1}{2} f_{0} 
\frac {\partial ^{2}}{\partial \xi ^{2}} (f_{0}^{2}) \right] .
\label{calf1}
\ee
The first two equations in (\ref{pert1}) can be integrated straight
away to give
\be
\delta _{1}(r)=0, \qquad
q_{1}(r) = K_{1}-\frac{3C_{1}}{4r} ,
\label{metric1}
\ee
where $K_{1}$ is an arbitrary constant and we have fixed the constant
arising in $\delta _{1}$ to be zero so that the geometry has the correct
asymptotic form at infinity.
The constant $K_{1}$ will be fixed by the requirement that there
is a regular event horizon at $r=r_{h}$.
This means that $q_{1}=\frac {r_{h}}{2}$, $q_{i}=0$ for $i\ge 2$
at $r_{h}$.  
The form (\ref{metric1}) corresponds to a Reissner-Nordstr\"om-AdS
geometry, as would have been expected.
We will therefore need to go to the next order in the metric functions
to get solutions with hair.
Integrating the equation for the Yang-Mills field in (\ref{pert1}) gives
\be
\frac {\partial f_{1}}{\partial r} =
\frac {1}{r^{3}-r_{h}^{3}}({\cal {C}}(\xi )r-{\cal {F}}_{1}(\xi )) ,
\ee
where ${\cal {C}}$ is an arbitrary function.
We require $f_{1}$ and all its derivatives to be regular everywhere
outside and on the event horizon, which fixes ${\cal {C}}$ as:
\be
{\cal {C}} r_{h}={\cal {F}}_{1} .
\ee
Integrating again then yields
\be
f_{1}(r,\xi )= {\cal {F}}_{2}(\xi )+
\frac {2{\cal {F}}_{1}}{r_{h}^{2}{\sqrt {3}}} 
\tan ^{-1} \left( \frac {2r+r_{h}}{r_{h}{\sqrt {3}}} \right) ,
\label{fpert1}
\ee
with another arbitrary function ${\cal {F}}_{2}(\xi )$.
Note that from (\ref{calf1}), ${\cal {F}}_{1}$ is of the form 
(\ref{fform}), so if we choose ${\cal {F}}_{2}$ to also be written as
(\ref{fform}), then the whole gauge field will have the structure 
required.
The function $f_{1}$ above consists of genuine $SU(\infty )$ gauge
field hair.
It contains two arbitrary functions of $\xi $ (subject to the restrictions
just discussed), which require an infinite number of constants 
in order to specify them (which correspond to the coefficients in the
expansion (\ref{fform})).

\begin{figure}
\begin{center}
\includegraphics[width=9cm,angle=270]{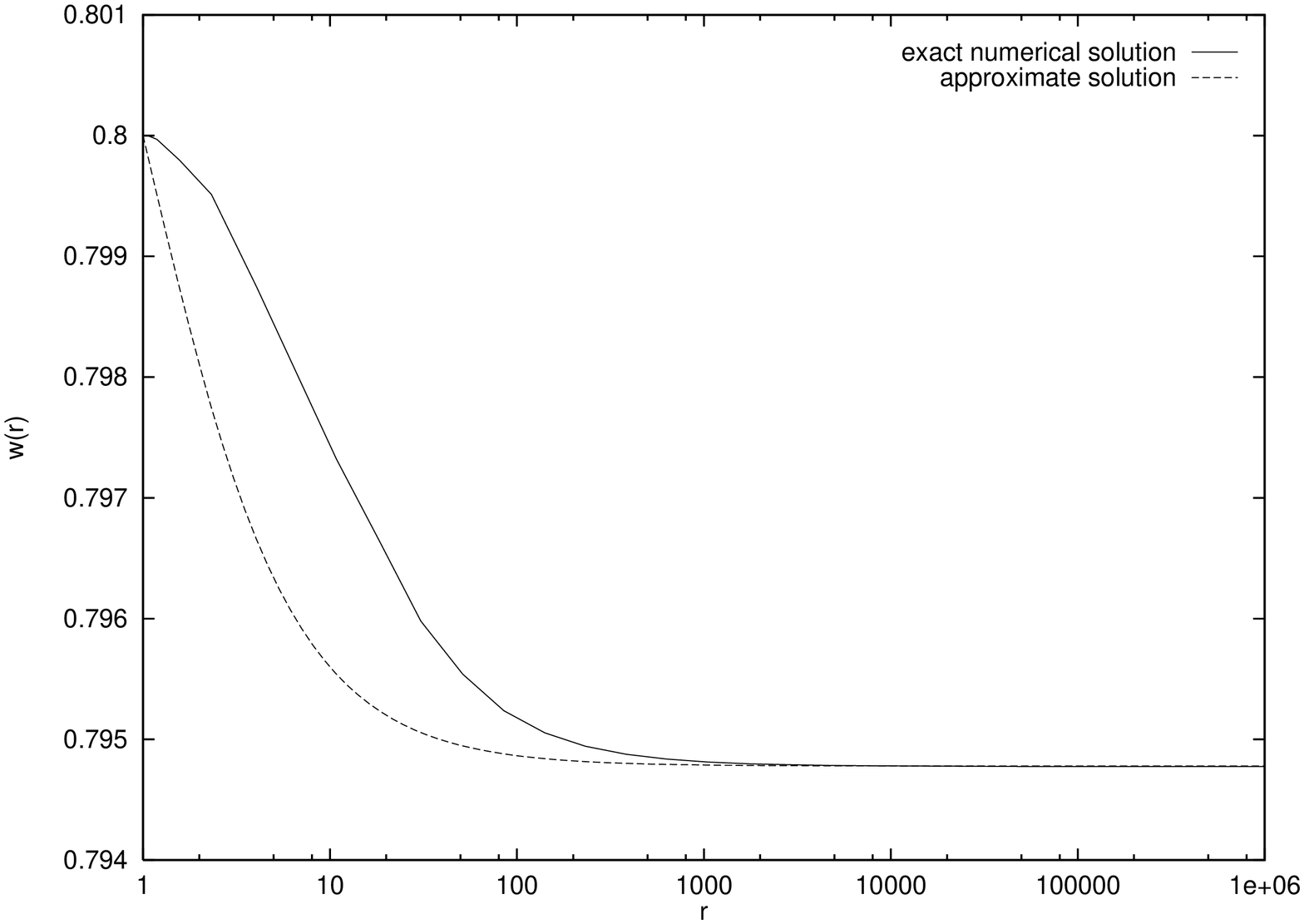}
\end{center}
\caption{Comparison of the exact, numerical solution for 
the gauge field with
gauge group $SU(2)$, where $\Lambda =-100$ and $w_{h}=0.8$
and the approximate solution given by (\ref{fpert1}).
The approximate solution has the correct qualitative
features, and simply converges a little too quickly.}
\label{check}
\end{figure}

\pr
We now want to find the effect on the geometry of the gauge field hair
(\ref{fpert1}).
To second order, the Einstein equations give the following equations
for the metric perturbations:
\bea
\delta _{2}'(r) & = & 
\frac {3C_{2}}{2rr_{h}^{2}} \left( r^{2}+rr_{h}+r_{h}^{2} 
\right) ^{-2}
\nn
\\
q_{2}'(r) & = &
\frac {C_{2}}{4rr_{h}^{2}} \frac {r_{h}-r}{r^{2}+rr_{h}+r_{h}^{2}}
+\frac {3C_{3}}{r^{2}} +
\frac {2{\sqrt {3}} C_{4}}{r^{2}r_{h}^{2}} 
\tan ^{-1} \left( \frac {2r+r_{h}}{r_{h}{\sqrt {3}}} \right)
\eea
where we have defined new constants $C_{2}$--$C_{4}$ by
\bea
C_{2} & = & \int _{\xi =-1}^{1} {\cal {F}}_{1}^{2}(\xi ) \, d\xi 
\nn
\\
C_{3} & = & \int _{\xi =-1}^{1} \left(
\xi +\frac {1}{2} \frac {\partial }{\partial \xi } (f_{0}^{2})
\right) \frac {\partial }{\partial \xi } 
(f_{0}{\cal {F}}_{2} ) \, d\xi 
\nn
\\
C_{4} & = & 
\int _{\xi =-1}^{1} \left( 
\xi +\frac {1}{2} \frac {\partial }{\partial \xi } (f_{0}^{2})
\right) \frac {\partial }{\partial \xi } 
(f_{0}{\cal {F}}_{1} ) \, d\xi .
\eea
Integrating then gives
\bea
\delta _{2}(r) & = &  
\frac {3C_{2}}{4r_{h}^{6}} \left[
\log \left( \frac {r^{2}}{r^{2}+rr_{h}+r_{h}^{2}} \right)
+\frac {2r_{h}(r_{h}-r)}{3(r^{2}+rr_{h}+r_{h}^{2})} 
-\frac {10}{3{\sqrt {3}}} \tan ^{-1} \left(
\frac {2r+r_{h}}{r_{h}{\sqrt {3}}} \right) \right]
+ K_{2}
\nn
\\
q_{2}(r) & = & 
\frac {1}{8r_{h}^{3}} \left( C_{2}+12 C_{4} \right)
\log \left( \frac {r^{2}}{r^{2}+rr_{h}+r_{h}^{2}} \right)
-\frac {3C_{3}}{r} 
\nn \\
& & 
-
\frac {1}{4r_{h}^{3}} \left(
{\sqrt {3}}C_{2} +12 C_{4} +\frac {24C_{4}r_{h}}{r} \right)
\tan ^{-1} \left( \frac {2r+r_{h}}{r_{h}{\sqrt {3}}} \right)
+K_{3}
\label{mpert2}
\eea
where $K_{2}$ and $K_{3}$ are arbitrary constants.
The constant $K_{2}$ is fixed by the requirement that 
$\delta _{2}\rightarrow 0$ as $r\rightarrow \infty $ to be
\be
K_{2}=\frac {5\pi C_{2}}{4 {\sqrt {3}}r_{h}^{6}} ,
\ee
whilst the constant $K_{3}$ will be fixed by 
the requirement that $q_{2}=0$ at $r=r_{h}$ so that we have
a regular event horizon.
We now have a hairy black hole metric, caused by the coupling
of the geometry to the $SU(\infty )$ Yang-Mills field.
In figure \ref{check} we compare the approximate solution 
(\ref{fpert1}), 
 with the exact numerical solution in the $SU(2)$
case, where $f$ has the form (\ref{fsu2}).
We use parameters  $\Lambda =-100$ and $w (r_{h})=0.8$.
It can be seen that the approximate solution converges a little too
quickly, but otherwise has the same qualitative features as the 
exact solution in this case.
The approximation would become better if  we used more terms
in the expansion (\ref{asympt}), or if $|\Lambda |$ was larger.

\pr
At this stage, a comment is in order concerning
the nature of the hair we have found.
In particular, we wish to stress that the hair is definitely 
infinite in quantity, and that an infinite number of 
parameters is required in order to describe it.
As mentioned above, the gauge field requires an infinite number of 
constants for it to be specified in this approximation, since 
two arbitrary functions of $\xi $ are involved.
Each subsequent term in the expansion in $\epsilon $ will introduce a 
further function of $\xi $ into the gauge field, and hence a further
infinite number of quantities.
Suppose, however, that we wish to fix the behaviour of the gauge
field at, say, the event horizon (this was the procedure used
to fix the constants in order to obtain figure \ref{check}, and 
is more convenient computationally than fixing quantities at infinity). 
This would require an infinite number of quantities to specify, say
$f_{0}(\xi )$. 
The subsequent functions of $\xi $ would be fixed by requiring that
$f_{i}(r,\xi )=0$ at $r=r_{h} $ for $i>0$, 
for example this would mean that
\be
{\cal {F}}_{2} (\xi ) = 
\frac {2\pi {\sqrt {3}} {\cal {F}}_{1}(\xi )}{r_{h}^{2}}.
\ee
The result, either way, is that an infinite number of parameters 
are needed to specify the gauge field function.
Next we turn to the metric functions.
These do not depend on $\xi $, but do contain constants constructed
from the gauge field function by integrating over $\xi $.
Each term in the perturbation expansion of the metric functions
involves at least one new constant constructed from the gauge field
function (which cannot be determined from previous constants) as well
as a constant of integration.
The constants of integration can easily be fixed by the
boundary conditions on the metric
i.e. by setting $\delta _{i}\rightarrow 0$ as $r\rightarrow
\infty $ for $i>0$ so that the geometry has the correct behaviour
at infinity, and setting $q_{i}=0$ at $r=r_{h}$ for $i\ge 2$ so
that there is a regular event horizon at $r=r_{h}$.
However, an infinite number of different constants, coming from
the gauge field function, will be required if we are to specify
the metric to all orders in $\epsilon $.
Thus, while it appears that we only need a finite number of constants
to construct the metric to each order in perturbation theory, the exact
metric will need an infinite number of constants, so that we do have
infinite amounts of hair.

\subsection{Solutions with a small gauge field}

\pr
In the previous subsection, we performed an asymptotic expansion
for very large $|\Lambda |$, and found analytic solutions 
representing black holes dressed with infinite amounts of 
gauge field hair.
We now turn to another situation in which we can find solutions
easily, and which is of direct relevance to 
holography and the
AdS/CFT correspondence~\cite{maldacena,witten} 
(which partly motivates our consideration
of black holes in AdS).
The holographic principle (namely that dynamics of gauge fields
in the bulk of AdS is determined by what is happening on the 
boundary) relies on a theorem of Graham and Lee~\cite{graham} 
which applies only to gauge fields sufficiently close to zero
\cite{witten}.
We stated previously that if the gauge field $f\equiv 0$,
then the geometry is that of non-extremal
Reissner-Nordstr\"om-anti-de 
Sitter (RN AdS) space, with charge $Q=1$.
Therefore in this section we are in effect considering
perturbations about this configuration.

\begin{figure}
\begin{center}
\includegraphics[width=9cm,angle=270]{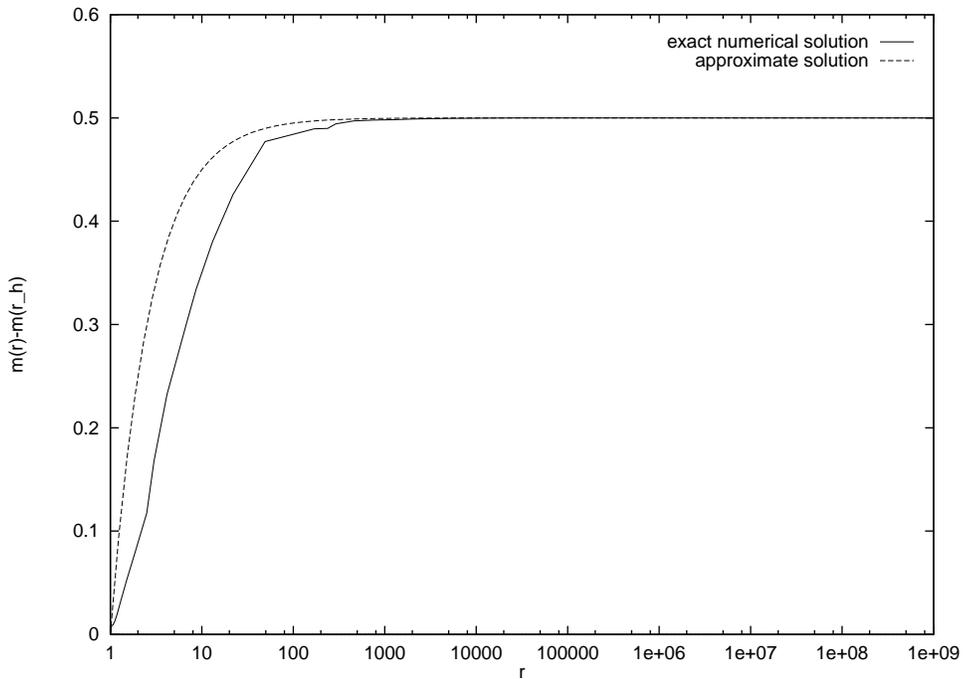}
\end{center}
\caption{Comparison of the exact, numerical solution
for the metric function $m(r)-m(r_{h})$
with gauge group $SU(2)$, where $\Lambda =-0.001$
and $w(r_{h})=0.1$ with the RN AdS solution, which is
an approximation when the gauge field is small.
From the figure it can be seen that the approximation 
is good in this case, even though the gauge field
is not particularly small.}
\label{inf}
\end{figure}

\pr
Let $\varepsilon $ be a small parameter such that
$\varepsilon \ll 1$ and $|f(r,\xi )|\le \varepsilon $
for all $r$, so that the magnitude of the gauge field is
uniformly bounded and is close to zero.
The parameter $\varepsilon $ determines the 
magnitude of the gauge field (no matter what the size of the 
cosmological constant $\Lambda$) 
and should  not be confused with $\epsilon = 1/\Lambda$ 
of the previous section where $|\Lambda| \gg 1$.  
This condition of uniform boundedness is rather strong,
as we shall see shortly, and the negative cosmological
constant is crucial if this condition is to be met.
We then consider the following asymptotic expansions of the
field variables as $\varepsilon \rightarrow 0$:
\bea
f(r,\xi ) & = & \varepsilon \left[
f_{0}(r,\xi ) +\varepsilon f_{1}(r,\xi )
+\varepsilon ^{2} f_{2}(r,\xi ) 
+\ldots \right]
\nn \\
m(r) & = & m_{0}(r)+\varepsilon m_{1}(r)
+\varepsilon ^{2} m_{2}(r)
+\ldots 
\nn \\
\delta (r) & = & \varepsilon \delta _{1}(r)
+\varepsilon ^{2} \delta _{2}(r) +\ldots .
\eea
These expansions are inserted into the field equations
(\ref{einstein},\ref{YMfinal},\ref{gravfinal}), as a consequence
of which $f_{1}$, $m_{1}$ and $\delta _{1}$ all
vanish identically.
In addition, we have, to lowest order, the RN AdS geometry, so that
\be
m_{0}(r)=\frac {1}{2r_{h}}+\frac {r_{h}}{2} 
-\frac {\Lambda r_{h}^{3}}{6} -\frac {1}{2r} .
\label{mzero}
\ee
In fact this is a good approximation to the exact metric
function $m$, as can be seen from figure \ref{inf},
where we compare (\ref{mzero}) with the exact numerical solution for the
$SU(2)$ case with $\Lambda =-0.001$ and $w(r_{h})=0.1$.
Again, the approximate solution has the correct qualitative
features and just converges a little too quickly.

\pr
The Yang-Mills equation, to lowest order, tells us that
the $\xi $ dependence of $f_{0}$ is arbitrary (subject to
(\ref{fform})), but its $r$
dependence is governed by the following differential equation:
\be
r^{2} \left( 1- \frac {2m_{0}}{r} -\frac {\Lambda r^{2}}{3}
\right) 
\frac {\partial ^{2}f_{0}}{\partial r^{2}}
+\left( 2m_{0} -\frac {2\Lambda r^{3}}{3} 
-\frac {1}{r} \right)
\frac {\partial f_{0}}{\partial r} +f_{0} =0.
\ee
This is exactly the same equation satisfied by very small
$SU(2)$ gauge fields, and in this case we can see that
the gauge field effectively becomes an infinite number of
$SU(2)$ degrees of freedom. 
The interaction between the degrees of freedom does not 
come into play in this approximation because the gauge field
is very small (and so we have ignored terms involving 
$f^{2}$).

\pr
We can now use our knowledge of $SU(2)$ EYM theory in AdS
\cite{eliz}
to comment on what will happen in the $SU(\infty )$ case.
From \cite{eliz}, there are solutions of $SU(2)$ EYM theory
in which the gauge field is uniformly bounded and arbitrarily small.
This is very different to the situation when $\Lambda =0$, when
$w\rightarrow \pm 1$ as $r\rightarrow \infty $, so that any
gauge field which is very small close to the event horizon, 
must become order unity at infinity.
In this case our expansion in terms of $\varepsilon $ is not 
uniformly valid.
For $\Lambda >0$, it is possible to have gauge fields which are
uniformly arbitrarily small, however, in that case, the 
set of parameters $w(r_{h})$ which lead to regular black hole
solutions consists of discrete points \cite{volkov}.
When the cosmological constant is negative, we have regular
black hole solutions for a continuous interval of 
values of $w(r_{h})$ containing zero, although the size
of this interval shrinks to zero as $\Lambda \rightarrow
0_{-}$  \cite{eliz}.
Therefore, for any negative cosmological constant (no matter 
how small or large), we can specify any 
sufficiently small function $f(\xi )$ 
satisfying the ansatz (\ref{fform}) to be the gauge field function
at the event horizon, and then we have infinite gauge field
hair which remains arbitrarily small at all distances away from 
the black hole horizon, including up to the boundary of AdS.
It is for this reason that we view the $\Lambda \rightarrow 0_-$  
model as a ``regulator'' model for $SU(\infty)$-hair 
on asymptotically flat space-times with zero-cosmological constant. 
A zero-cosmological-constant situation might be desirable
from both theoretical and phenomenological points of view, 
e.g. it may be the result of a space-time symmetry like supersymmetry. 
Notably, in the two-dimensional stringy analogue of the $W$-hair~\cite{emn},
the ground-state energy of the two-dimensional string 
has been preserved under the action of the $W_\infty$ 
symmetries~\cite{ring} 
and this could be viewed as the analogue of a 
zero-cosmological constant situation.

\section{Conclusions} 

\pr
In this work we have formulated 
the field equations for spherically-symmetric 
EYM black holes
in $SU(\infty)$ gauge theory.
Their study was motivated by some considerations 
in string theory, the holographic principle and the so-called
AdS/CFT correspondence.
We have solved these equations
both numerically and analytically in some cases. 
The introduction of a negative cosmological 
constant (no matter how small) proved extremely
useful in simplifying the equations by means 
of appropriate perturbative expansions. 
Specifically, we have obtained   
solutions for the cases: (i) $\Lambda < 0$ with
$|\Lambda | \gg 1$, 
and finite gauge fields, 
and (ii) $\Lambda  < 0$ arbitrary, 
and infinitesimally small gauge fields. 
The latter case exhibits holographic properties
of the gauge-field hair and may be relevant
for issues related to information loss. 
\pr
However, as stressed in the introduction, 
the issue of quantum coherence is still far from 
being understood. 
The above holographic properties pertain to 
classical field configurations, and 
it is not clear that they survive a full quantum treatment,
where space-time fluctuations are taken into account. 
In ref. \cite{emnw} we have presented an approach to 
quantify entropy production 
during an evaporation process
for the case of an $SU(2)$ EYM 
black hole, coupled to scalar matter~\cite{mw}.
This can be done by 
identifying the target-space temporal evolution parameter, `time' $t$,
with the logarithm of the brick-wall scale, interpreted as
a Renormalization-Group scale that regularizes 
ultraviolet (short-distance) infinities in the coupled matter/black
hole system.
One then 
concentrates on the logarithmic divergences
in the expression for the entropy of the scalar 
field.
Such divergences are extra divergences that cannot 
be absorbed in a standard renormalization of Newton's 
or other field-theory coupling constants (which for instance
yield a regularized Bekenstein-Hawking area law~\cite{mw}). 
In that sense, such terms have been interpreted
in \cite{emnw} as entropy production. 
\pr
If the holographic properties of the 
$SU(N \rightarrow \infty )$ EYM 
AdS background
were to be extended to the quantum case, 
such logarithmic divergences should have been suppressed
by inverse powers of $N \rightarrow \infty$. 
However, it can be easily seen that this is not the case
for a minimally-coupled scalar field neutral under the gauge symmetry.
Indeed, the geometries we consider are small perturbations
near RN AdS black-hole space times, and therefore the entropy
of a scalar field interacting with this geometry
is very nearly the same as the RN AdS case,
which is known to be non-trivial~\cite{entropy}.
However, the case of a scalar field coupled to the gauge degrees of freedom,
which exhibit the holographic properties, may completely change the picture.
We hope to return to this issue in a future work. 
Nevertheless, we believe that the results presented in this work
are of sufficient interest to motivate further studies in 
the above direction. 

\section*{Acknowledgements} 

It is a pleasure to thank John Ellis for useful discussions and 
for suggesting the title~!
The work of N.E.M. is partially supported by P.P.A.R.C. (U.K.). 
E.W. wishes to thank Oriel College (Oxford) for financial support and
the Department of Physics, University of Newcastle (Newcastle-upon-Tyne)
for hospitality.

\newpage

\end{document}